\documentclass[aps,prl,reprint,groupedaddress,showpacs]{revtex4-1}
\usepackage{graphicx}

\begin{document}


\title{Frequency Comb Velocity-Modulation Spectroscopy}


\author{Laura C. Sinclair}
\email[]{sinclail@jila.colorado.edu}
\author{Kevin C. Cossel}
\author{Tyler Coffey}
\author{Jun Ye}
\author{Eric A. Cornell}
\affiliation{JILA, National Institute of Standards and Technology and University of Colorado
Department of Physics, University of Colorado, Boulder, CO 80309-0440, USA}

\date{\today}

\begin{abstract}
We have demonstrated a new technique that provides massively parallel comb spectroscopy sensitive specifically to ions through the combination of cavity-enhanced direct frequency comb spectroscopy with velocity modulation spectroscopy.  Using this novel system, we have measured electronic transitions of HfF$^{+}$ and achieved a fractional absorption sensitivity of 3 x $10^{-7}$ recorded over 1500 simultaneous channels spanning 150 cm$^{-1}$ around 800 nm with an absolute frequency accuracy of 30 MHz (0.001 cm$^{-1}$). A fully sampled spectrum consisting of interleaved measurements is acquired in 30 minutes.
\end{abstract}


\maketitle

\setlength{\parskip}{0pt}
\linespread{0.75}
\setlength{\textfloatsep}{2pt}
\setlength{\dbltextfloatsep}{2pt}

Broad bandwidth, precision spectroscopy of molecular ions is of interest for applications in precision measurement, astrochemistry, and physical chemistry.  While molecular ions such as HfF$^{+}$ and ThF$^{+}$ may be ideal for a highly sensitive search for the electron electric dipole moment \cite{Meyer2006}, there is very rarely any experimental data on the electronic structure of these ions, and the large theoretical uncertainties (thousands of wavenumbers) \cite{Petrov2007} necessitate broad survey spectroscopy to understand and assign the molecular energy level structure. On a different front, precision spectroscopy of molecular ions such as H$_{3}^{+}$ \cite{Morong2009} and CH$_{5}^{+}$ \cite{Huang2006} can provide a test of ab initio theory and shed new insights into few-body quantum dynamics, respectively. In astrochemistry, numerous ions and radicals have been discovered in the interstellar medium and many more species have been hypothesized \cite{Klemperer2010,Snow2008}. It is believed that the source of the unidentified diffuse interstellar bands, currently the longest standing question in astronomical spectroscopy, is molecular (possibly ionic), but accurate identification will require matching the spectra of candidate species to astronomical observations over wide and often times disparate spectral windows \cite{Sarre2006,Maier2011}.  Additionally, understanding the chemistry of molecular clouds in circumstellar gases has been enabled by the identification of molecular ions in the laboratory \cite{Halfen2007}.  Efficient survey spectroscopy of molecular ions is thus of interest to a wide range of applications.

The application of optical frequency combs to a variety of spectroscopic techniques has resulted in ultrahigh sensitivity systems with simultaneous high resolution, absolute frequency accuracy, and spectral bandwidth, all with dramatic reduction in data acquisition times \cite{Marian2004,Keilmann2004,Thorpe2006,Diddams2007,Coddington2008,Bernhardt2010,Adler2010}.  Prior to this Letter, no techniques have been capable of ion-specific high-sensitivity modulation spectroscopy on every parallel detection channel over a broad spectral range.  The integration of velocity modulation spectroscopy \cite{Stephenson2005,Gudeman} and cavity-enhanced direct frequency comb spectroscopy \cite{Thorpe2006, Adler2010} provides a way to achieve very broad spectral bandwidth ion-sensitive spectroscopy without sacrificing high-resolution or high-sensitivity.

\begin{figure}
\includegraphics{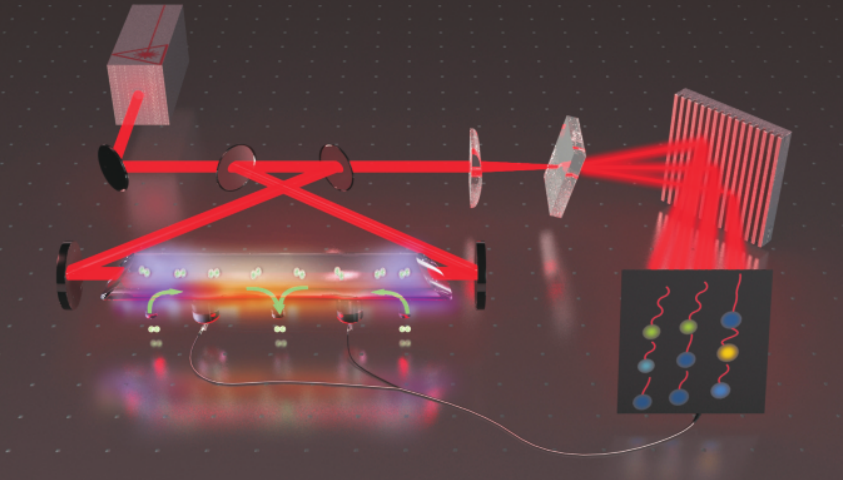}
\caption{\label{figure1}Qualitative sketch of a frequency comb velocity-modulation spectroscopy system.  Light from the frequency comb is coupled into a bow-tie ring cavity containing an AC discharge tube.  The AC discharge modulates the ions' Doppler shift and thus the ion absorption signal.  The transmitted light is dispersed by a VIPA etalon and a grating to produce a 2-dimensional image in which every comb tooth is resolved and detected.  A fast demodulation camera performs lock-in detection of each comb tooth at the discharge modulation frequency to recover ion-specific absorption signals across each comb mode. (Figure courtesy B. Baxley(JILA).)}
\end{figure}

High resolution spectroscopy of molecular ions is typically performed using velocity-modulation spectroscopy (VMS), first demonstrated in Saykally's group \cite{Stephenson2005,Gudeman}.  Velocity-modulation spectroscopy achieves high sensitivity via lock-in detection of the absorption signal from a laser which passes through a sample of ions with a modulated Doppler shift due to an alternating current discharge \cite{Gudeman}.  Although recent experiments have pushed to higher sensitivity \cite{Morong2009,Gottfried2003} as well as higher accuracy \cite{Mills2010,Siller2010}, these systems all rely on slow scans of continuous wave (CW) lasers.  In addition, broad spectral bandwidth has been achieved via emission spectroscopy \cite{Martin1990,Picque1999}, but without the methods of enhancing sensitivity of the current experiment.  Comb-based spectroscopic techniques provide broad simultaneous bandwidth with high resolution and high sensitivity \cite{Adler2010}.   We find that cavity-enhanced direct frequency comb spectroscopy \cite{Thorpe2006,Adler2010} is an ideal candidate to integrate large instantaneous bandwidth with high resolution VMS by enabling lock-in detection simultaneously on every comb tooth.

To integrate these two powerful spectroscopic techniques, several key features are needed, as sketched in Fig. \ref{figure1}.   An alternating-current discharge tube for the production and modulation of molecular ions must be placed within an optical enhancement cavity, with the cavity length actively stabilized such that the frequency comb light can be coupled into the cavity.  The cavity needs a ring geometry to provide a single direction of propagation of the light relative to the instantaneous motion of the ions, which enables lock-in detection at the discharge frequency (10 kHz).  For frequency accuracy that is not limited by the imaging system resolution, the imaging system must be able to resolve each individual comb mode.  Finally, the detector used must be able to demodulate the intensity of each comb mode at 10 kHz to recover the absorption signal.

\begin{figure}
\includegraphics{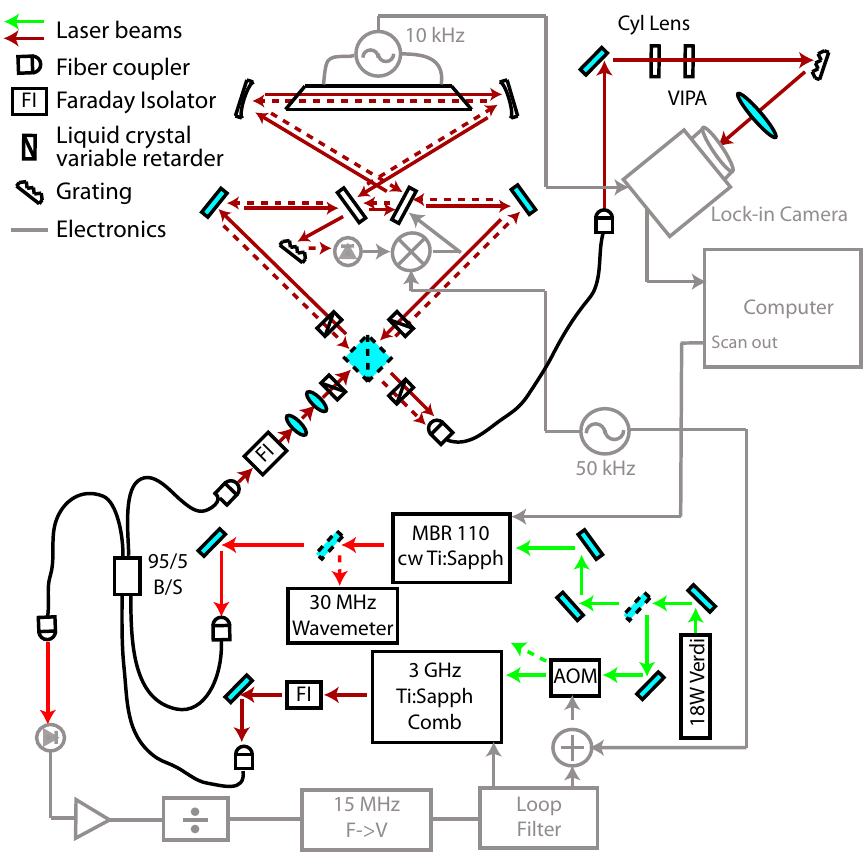}
\caption{\label{figure2} Schematic of the system. An 18 W Coherent Verdi pumps both the CW Ti:Sapphire laser, used as a frequency reference, and the 3 GHz repetition rate Ti:Sapphire comb.  The CW laser is read by a wavemeter with an accuracy of 30 MHz.  The comb is referenced to the CW laser via a beatnote with a single comb tooth.  Feedback to the comb's repetition rate is achieved through the pump power (high bandwidth and small range) and through the comb cavity length (low bandwidth and large range).  95$\%$ of the comb light and 5$\%$ of the CW light are combined in a fiber which is coupled to the enhancement cavity.  Liquid crystal retarders and a polarizing beam splitter are used to select the direction of light propagation around the enhancement cavity while maintaining all locks.  The enhancement cavity is locked to the comb.  Cavity transmitted light is dispersed by the 2-dimensional imaging system, qualitatively sketched in Fig. \ref{figure1}, and recorded by the lock-in camera, which demodulates each pixel at the discharge modulation frequency.  The CW light serves as a reference marker on the lock-in camera image.}
\end{figure}

Figure \ref{figure2} provides a more detailed layout of the system.  A 2.5 meter long bow-tie enhancement-cavity with a finesse of 100 surrounds a 1 meter long discharge tube, which is driven at 10 kHz with $\approx$ 250 mA$_{pk-pk}$.  The discharge tube contains HfF$_{4}$, which is heated to 800 K, and helium buffer gas is flowed through the discharge tube such that the total pressure is 5 torr.  A Gigaoptics 3 GHz repetition rate Ti:saphhire comb is coupled into the 120 MHz free-spectral-range enhancement-cavity traveling either clockwise or counterclockwise around the bow-tie cavity, with every 25$^{th}$ cavity resonance matched to a comb mode.  Spectral filtering of the cavity due to group velocity dispersion (GVD) limits the transmitted bandwidth to 300 cm$^{-1}$.  The windows on the discharge tube are at Brewster's angle to minimize intra-cavity loss and are thin to reduce GVD.

The cavity-transmitted light is coupled via fiber to a two-dimensional lock-in imaging system.  The 1 GHz resolution crossed-dispersion system consisting of a VIPA etalon \cite{Xiao2004,Shirasaki1996} and a grating is used to resolve each individual comb mode \cite{Diddams2007,Thorpe2008}.   We used a Heliotis C2 ``smart pixel'' lock-in camera \cite{Beer2005,Bourquin2003}, which uses a combination of CCD and CMOS technology to demodulate and read out each pixel at the discharge modulation frequency.  By rapidly switching between clockwise and counterclockwise beam propagation via polarization control through the ring cavity and subtracting the two images, we are able to reduce the impact of drifts in the gain of each individual pixel on long time scales as well as remove other common mode noise and further increase the rejection of neutral molecular signals while retaining the ion absorption signals.

A CW Ti:sapphire laser serves as a frequency reference for stabilizing the comb and provides a marker on the two-dimensional spectral image. The CW laser is locked to its internal reference cavity, and feedback to the comb's repetition rate locks a comb tooth to the CW laser.  The repetition rate is controlled by acting on both the comb laser cavity length (slow feedback via a piezoelectric transducer) and the pump power (fast feedback via an acousto-optic modulator) \cite{Holman2003,Heinecke2009},  while the offset frequency of the comb is left free-running.  The enhancement cavity is then locked to the comb via a dither lock using the spectrally filtered reflection from the cavity.  We found it critical to feed forward a 10 kHz compensation voltage onto the lock of the enhancement cavity to compensate for the 10 kHz pickup from the discharge, an effect seen in other cavity-based velocity-modulation experiments \cite{Siller2010} as well.

\begin{figure*}
\includegraphics{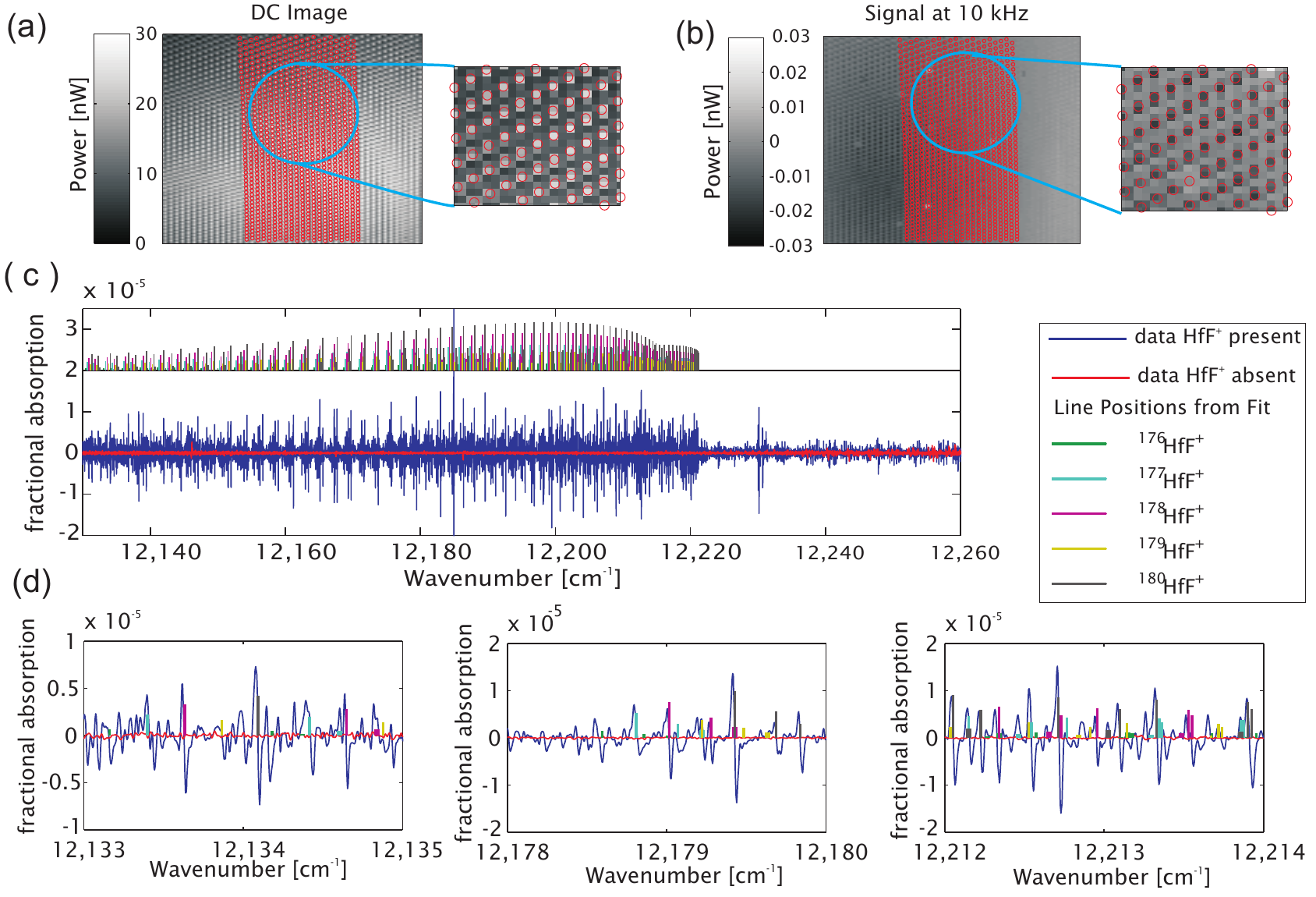}
\caption{\label{figure3} Data Analysis.(a) The DC intensity is shown as the initial 2-dimensional image, which is then fit
to obtain the comb mode positions as shown in red circles over part of the image. In (b) the signal
at 10 kHz (with only the discharge modulation) is shown for one image. The red circles on the 2-
dimensional image are the positions of the comb modes determined from the global fit to the DC
intensity in (a).  The resulting processed spectrum with HfF$^{+}$ present (blue) and absent (red) from all the images (each
processed similarly but with stepped CW laser frequency) is shown in (c) along with the line positions from a fit to one of the bands visible (offset for clarity) for all five isotopes of hafnium.  Much of the additional unlabeled structure is due to overlapping rotational bands. The full recorded spectrum was acquired in 30 images with 1500 simultaneous channels in 30 minutes, with a sensitivity of 3 x 10$^{-7}$ fractional absorption per channel. The zoomed-in panels (d) show three different portions of the spectrum with the line positions from the fit.  (Fits to the HfF$^{+}$ bands will be discussed in a later publication.)}
\end{figure*}

Figure \ref{figure3} shows our spectrum generation process. To collect a spectrum with sufficiently dense sample points, a signal image (Fig. \ref{figure3}(b)) is taken at each of 30 steps of a 3 GHz scan of the CW laser while maintaining all locks.  Each image contains absorption information at every comb tooth, 3 GHz spacing, across approximately 150 cm$^{-1}$ with the spectral bandwidth limited by the size of the detector array.  In addition, to remove the comb intensity noise and drift, a separate reference image extracts the DC intensity (Fig.\ref{figure3}(a)) by means of a calibrated 10 kHz amplitude modulation applied to the light transmitted through the cavity.  A global fit to the DC intensity image yields the spatial location of each comb tooth (red circles Fig. \ref{figure3}(a)).  Based on the fit location of each comb spot on the grid formed by the pixels, one to four pixels are summed over to capture the signal at 10 kHz and DC intensity for that comb spot.  After summation, the signal at 10 kHz is normalized by the DC intensity of the corresponding spot, and the difference is taken between the data for clockwise and counterclockwise beam propagation.

Residual 10 kHz noise from the discharge is present on the cavity transmitted light due to incomplete cancellation by the feed-forward mechanism resulting in baseline offsets.  These offsets arise from frequency-to-amplitude-noise conversion in the enhancement cavity.  Since the offsets have common values across neighboring comb modes, they are removed by subtracting the median value across a number of neighboring comb modes.  This method of rejecting common mode noise, inherent in this comb-based spectroscopy, is not available for conventional CW systems.

Frequency assignment of each comb tooth is based on the knowledge of the repetition rate of the comb and the CW laser frequency. The repetition rate is measured using a fast photodiode fed into a frequency counter, and the CW frequency is recorded using a calibrated wavemeter with an accuracy of 30 MHz.  The frequency of the comb mode that is locked to the CW laser is thus known, so the frequencies of all comb modes in an image are known precisely.  We detect about 1500 comb modes in a particular spectral acquisition. With the drift of the repetition rate due to changes in the carrier envelope offset frequency for a stabilized comb on the order of 100 Hz, the frequency uncertainty of any comb modes with respect to the mode that is locked to the CW laser is under 1 MHz.  By interleaving data from all the images, we can construct a densely sampled absorption spectrum with the approximately first derivative lineshapes typical of VMS as shown in Fig. \ref{figure3}(c).

In Fig. \ref{figure3}(c) and Fig. \ref{figure3}(d), we show part of the measured spectrum for previously unrecorded HfF$^{+}$ bands.  Three different HfF$^{+}$ bands have been identified with the rich isotope structure of hafnium resolved.  Molecular constants have been assigned from fits to the hundreds of lines recorded and will be discussed in a later publication.  The inherent broad spectral bandwidth of this technique has allowed for measurement of HfF$^{+}$ bands even given the large theoretical uncertainties in the energy level structure.

Our achieved sensitivity here is 3 x $10^{-7}$ single-pass fractional absorption for each measurement channel, with 1500 simultaneous channels covering 150 cm$^{-1}$, corresponding to 4 x 10$^{-8}$ Hz$^{-1/2}$ (spectral element)$^{-1/2}$ absorption sensitivity. We used 30 interleaved images to densely sample the whole spectrum in 30 minutes.  For other high-accuracy CW-laser velocity-modulation experiments using cavity-enhancement, the current sensitivity is $\approx$ $10^{-5}$ \cite{Siller2010} compared to our sensitivity of 3 x 10$^{-7}$.  We note that for a CW laser system to achieve the same level of sensitivity across the spectral bandwidth we cover in a single acquisition, it would need a sensitivity of 4 x 10$^{-8}$ Hz$^{-1/2}$ and the ability to scan across 150 cm$^{-1}$ continuously.  The technical noise in our current system, due to incomplete common-mode noise rejection, is a factor of four above the readout noise on the camera, which in turn lies above the shot-noise limit.  Modifications of the lock-in camera, which was designed for higher power applications, could also improve sensitivity.

Frequency comb modulation spectroscopy can be applied to both ionic and neutral molecules by either demodulating at the discharge modulation frequency (ion velocity modulation) or at twice the discharge modulation frequency (neutral-molecule production modulation).  Comb sources also enable efficient non-linear optical generation, which allows for spectral broadening and access to spectral regions ranging from the UV to the mid-IR \cite{Gambetta2008,Adler2009,Cossel2010}, but we note that the system is already capable of high-sensitivity, broadband survey spectroscopy of previously unmeasured transitions in HfF$^{+}$ and will soon be applied toward measurements of new transitions in ThF$^{+}$ for which the theoretical uncertainties are yet larger.

\begin{acknowledgments}
We thank F. Adler, W. Ames, R. Stutz and M.J. Thorpe for technical discussions and assistance. Funding for this work is provided by NIST, NSF, and AFOSR.
\end{acknowledgments}


%

\end{document}